\documentclass[final,5p,times]{elsarticle}

\usepackage{amssymb}
\usepackage{amsthm}
\usepackage{lineno}
\usepackage{lipsum}
\usepackage[dvipsnames,table,xcdraw]{xcolor}

\usepackage{url}
\usepackage{etoolbox}
\patchcmd{\emailauthor}{(#2)}{}{}{}
\patchcmd{\urlauthor}{(#2)}{}{}{}

\newcommand\rvwA[1]{\textcolor{black}{#1}}

\journal{Physics in Medicine and Biology}

\graphicspath{{Figures/}}

\begin{document}

\title{\Large \textbf{\rvwA{A novel range telescope concept for proton CT}}}
\author[a]{M.~Granado-Gonz\'alez\corref{cor1}}
\ead{mgg@hep.ph.bham.ac.uk}
\cortext[cor1]{Corresponding author}
\author[b]{C.~Jes\'us-Valls\corref{cor1}}
\ead{cjesus@ifae.es}
\author[b]{T.~Lux}
\author[a]{T.~Price}
\author[c]{F. Sánchez}

\address[a]{University of Birmingham, School of Physics and Astronomy, Edgbaston, Birmingham B15 2TT, UK}
\address[b]{Institut de F\'isica d’Altes Energies (IFAE) - The Barcelona Institute of Science and Technology (BIST), Campus UAB, 08193 Bellaterra (Barcelona), Spain}
\address[c]{ University of Geneva, Section de Physique, DPNC, 1205 Genève, Switzerland}

\author{}
	
\begin{abstract}
\small

Proton beam therapy can potentially offer improved treatment for cancers of the head and neck and in paediatric patients. There has been a sharp uptake of proton beam therapy in recent years as improved delivery techniques and patient benefits are observed. However, treatments are currently planned using conventional x-ray CT images due to the absence of devices able to perform high quality proton computed tomography (pCT) under realistic clinical conditions.\rvwA{A new plastic-scintillator-based range telescope concept, named ASTRA, is proposed here to measure the proton's energy loss in a pCT system. Simulations conducted using GEANT4 yield an expected energy resolution of 0.7\%. If calorimetric information is used the energy resolution could be further improved to about $0.5\%$.  In addition, the ability of ASTRA to track multiple protons simultaneously is presented. Due to its fast components, ASTRA is expected to reach unprecedented data collection rates, similar to $10^8$ protons/s. The performance of ASTRA has also been tested by simulating the imaging of phantoms. The results show excellent image contrast and relative stopping power reconstruction.}
\end{abstract}


\maketitle

\section{Introduction}
\noindent \rvwA{Protons deposit more energy as they slow down, and deliver a very localized dose close to their stopping point. Consequently, proton beam radiotherapy (PBR) is a potential alternative to the well established X-ray radiotherapy~\cite{Bryant963} to reduce the treatment toxicity to healthy tissue \cite{GABANI2019116,Suneja}. However, PBR still cannot be fully exploited as technical challenges remain to be overcame. To address treatment planning a high quality 3D map of the body's Relative Stopping Power (RSP) is needed.}
The state-of-the-art technique is to obtain a computed tomography (CT) of the target using X-rays and to map the image into RSP. \rvwA{This procedure introduces uncertainties of around 1.6 $\%$ (0.7$\%$)  for single (dual) energy CT~\cite{bar2018experimental}. Alternatively, a direct measurement of the RSP could be directly achieved by doing a proton CT (pCT). In addition to eliminate the systematic errors arising from mapping photon measurements into RSP, pCT could lead to shorter time scans, lower toxicity rates and would provide repeatability conditions enabling pre-treatment image verifications. Consequently, the development of a pCT system is of great interest. Several significant steps have been already achieved. For instance, the Loma-Linda/UCSC Phase-II scanner was able to produce images of a human-head size anatomy~\cite{6701142}, and to experimentally demonstrate good imaging capabilities paired with measurement rates similar to 1~MHz~\cite{7352382}. The PRaVDA collaboration introduced the possibility of using a full solid-state system for pCT, with good imaging results~\cite{ESPOSITO2018149}. More recently, the ProtonVDA LLC has developed a prototype aimed to operate at 10~MHz~\cite{dejongh2021fast}. Despite of the current progress, clinical devices able to perform pCT are not yet available. The limitations are often either the production costs, the achievable measurement rates, or the imaging resolutions.\\
In this article we propose a novel range telescope concept with the potential to overcome several of the current limitations and we discuss its expected performances based on existing data and Monte Carlo simulations. The article is organized as follows: Section 2 presents the design details for all the elements in the proposed pCT system; Section 3 discusses the simulation set-up and analysis methodology; Section 4 presents the performance results and Section 5 presents the conclusions.}
\section{pCT design}
\begin{figure}[h!]
	\centering
	\includegraphics[width=0.99\linewidth]{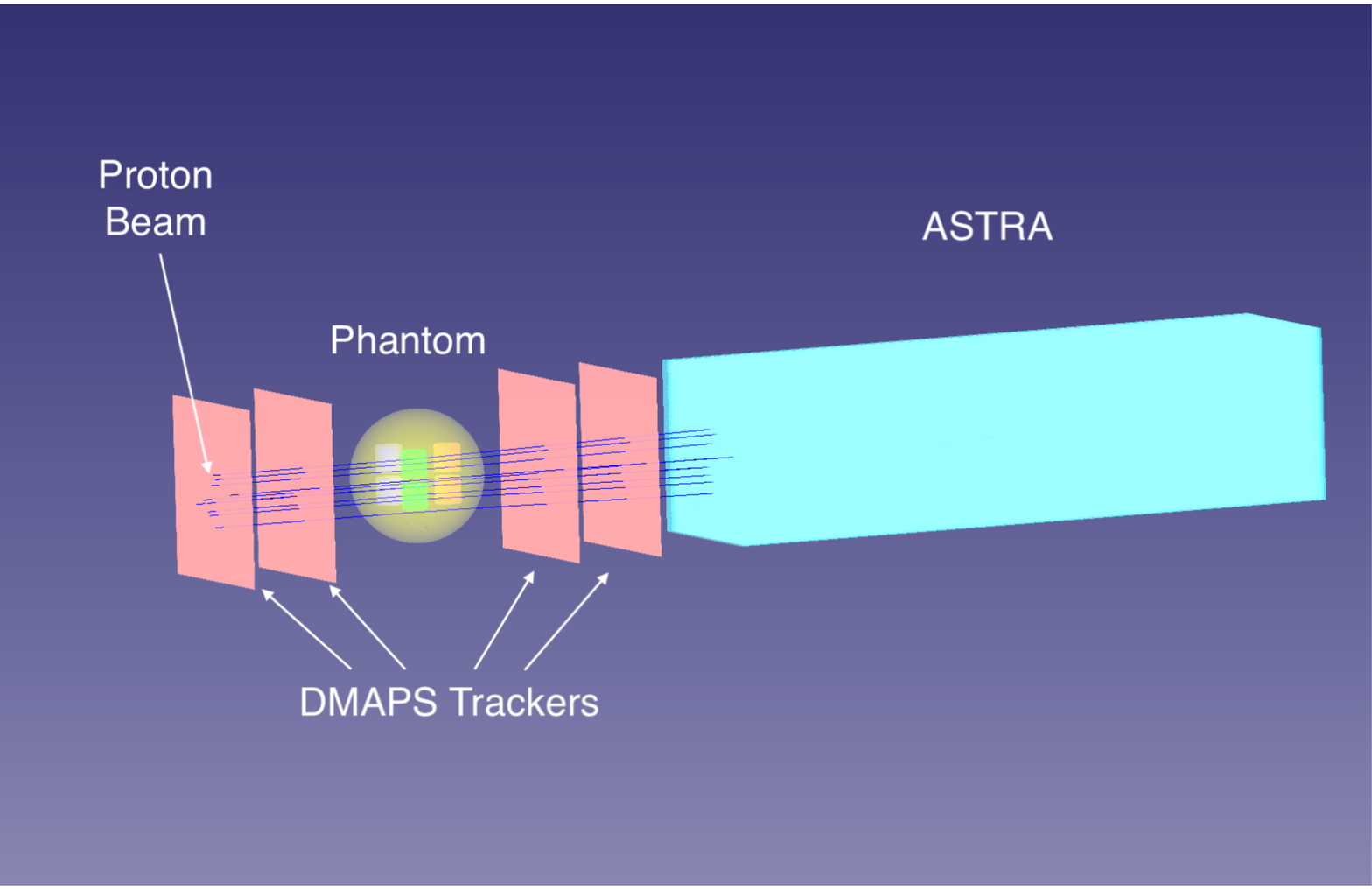}
	
	\caption{3D visualisation of the pCT system with 4 DMAPS layers, a spherical phantom with 6 cilindrical inserts in place and the ASTRA range telescope. In the image, ten protons (dark blue lines) are being measured.}
	\label{fig:sketch}
\end{figure}

A pCT system requires:
\begin{itemize}
    \item A \textit{position tracker}, able to reconstruct the proton trajectory within the body.
    \item An \textit{energy tagger}, able to reconstruct the proton energy.
\end{itemize}
\rvwA{In order to study the potential pCT performances of the proposed energy tagger, we envisioned a full pCT system}, sketched in Figure~\ref{fig:sketch}. \rvwA{It consists} in a position tracker made up of four Depleted Monolithic Active Pixel Sensors (DMAPS) placed in pairs either side of a phantom and A Super-Thin RAnge telescope (ASTRA) located downstream. 
\subsection{DMAPS-based tracker}
DMAPS have been developed for the high-luminosity upgrade of the inner tracker of the ATLAS detector~\cite{Pernegger:2017trh,Wang:2016dgi}. \rvwA{Thanks to the fully depletion of the bulk DMAPS have a fast response and fast frame rate of 40~MHz.}  Thus, together with the highly segmented pixel sizes, on the order of tens of microns, privide them with excellent properties to build proton trackers for pCT. In addition, their radiation hardness ensures a long life-time of the sensors even in highly active and high radiation environments~\cite{Terzo:2019efa} and the monolithic approach provides a compact solution without bump bonding reducing production and maintenance costs. \rvwA{Noticeably, similar pixel technologies, such as the ALICE pixel detector (ALPIDE), have also been considered for pCT~\cite{alme2020high}}.
\\The DMAPS-based tracker that we consider would consist of four identical DMAPS, organized in two sub-trackers, front and back, each formed by a pair of DMAPS separated by 50~mm. We consider the gap distance between the front and back trackers to be of 150~mm. We simulate the DMAPS as being similar to those in reference \cite{Neubuser:2020dvr}, 100~\textmu m thick with a shallow sensitive layer segmented forming an array of 2500x2500 silicon pixels of 40$\times$40~\textmu m$^2$ covering a total area of 10$\times$10~cm$^2$. Notice that such a device has does not yet exist but could be produced on a reasonable timescale. 

\subsection{ASTRA}
The ASTRA detector is a novel concept presented in this article, inspired by the geometry of the existing Fine Grained Detector (FGD) modules~\cite{Amaudruz:2012esa} in the ND280 detector of the Tokai-to -Kamioka (T2K) experiment~\cite{T2K:2011qtm} and by recent R\&D in plastic scintillator detectors, such as the time-of-flight panels~\cite{Betancourt:2017sex,Korzenev:2019kud} and the SuperFGD detector \cite{Sgalaberna:2017khy,Abe:2019whr} developed in the context of the ND280 detector upgrade~\cite{Abe:2019whr}.
\\The ASTRA detector will be a plastic-scintillator range telescope consisting of layers made up of thin polystyrene bars oriented in alternate axis, perpendicular to the proton beam. \rvwA{The exact plastic choice could be the EJ-200 plastic scintillator, which has a scintillation rise time of 0.9~ns, a decay time of 2.1~ns, and an attenuation length of 380 cm~\cite{eljen}}. Here, we consider \rvwA{a prototype size for ASTRA, consisting of} bars of 3$\times$3$\times$96~mm$^3$, arranged in groups of 32 bars per layer. This provides a cross-section of 9.6$\times$9.6~cm$^2$, well matching the area of the DMAPS tracker. \rvwA{If necessary, the cross-section of ASTRA could be easily scaled up by increasing the number of bars per layer. The length of ASTRA can be tuned to match the maximum beam energy, optimizing the production costs. Here, we simulated a length of 360~mm (120 layers), enough to stop protons of 240~MeV.} Following reference~\cite{Blondel:2020hml} each bar is simulated including an inactive polystyrene layer of 50~\textmu m, necessary to achieve good bar-to-bar optical separation. In practice, the bars will be manufactured by plastic extrusion, and the outer layer produced by etching the bar surface~\cite{Blondel:2020hml}. The bars would be readout by a SiPM directly coupled to the scintillator bulk, as in reference~\cite{Korzenev:2019kud}. \rvwA{In order to match the fast plastic response of the instrument, the fast output pulse shape from the Onsemi's MicroFJ SiPMs~\cite{onsemi} could be used, providing a full waveform in the span of few nanosecons. Suitable choices for the electronics already exist, such as those used in reference~\cite{Blondel:2020hml}, which provide a deadtime free readout at a 0.4~GHz sampling rate. Remarkably, an improved version of the electronics is under development with the goal to achieve 0.8~GHz. Under this specifications, the light of two consecutive protons separated by a time span equal or higher to 10~ns can be realistically expected to be well separated, and accordingly, ASTRA has the potential to reach an event rate equal or higher to 10$^8$ protons/s (100~MHz). Nonetheless, it must be noted that further reducing the proton's time gap to less than 10~ns with this same system could be possible, e.g. by doing a shape analysis of the the waveform, and deserves dedicated attention in the future.}\\
\rvwA{Concerning ASTRA'a geometry two main motivations drive its design. First, by using bars instead of layers the residual energy can be precisely reconstructed by range even if protons do not follow a perfectly straight trajectory. Second, if the beam has a typical spread comparable to the size of few of its bars, multiple protons can be tracked simultaneously when the time information is not enough to discriminate their trajectory. This has the advantage of further increasing the already high rate capability and reducing detector inefficiencies when the beam can not be perfectly controlled to deliver a single proton per time frame.\\}
\rvwA{Despite that other scintillator-based range telescopes have been proposed and tested in the past, e.g. see references~\cite{sadrozinski2013development,ulrich2021proton,7582123,bucciantonio2013development}, ASTRA will introduce many significant novelties to this field. In one hand, it has been designed to reach collection rates two orders of magnitude higher than previous technologies. In the other hand, all previous designs were based on layers, instead of bars, limiting the device intrinsic resolution. Finally, the novelty of coupling the SiPM directly to the scintillator bulk will eliminate the necessity of introducing dead material inside sensitive volume of the detector, such as wavelength shifting fibers.}

\section{Methodology}
The system as described in the previous section has been simulated using GEANT4~\cite{Agostinelli:2002hh} with the \texttt{QGSP$\_$BIC} physics list. 
The energy deposit in each of the DMAPS planes has been discretized in a list of fired pixels analogous to the real output. A threshold of 850 electrons, far from the signal's most probable value of $\sim$20000~electrons, has been used in order to minimize the noise coming from secondary electrons.
\\The energy deposit in ASTRA is discretized in a list of bar hits. In each bar, the energy deposit, is converted into a number of SiPM photo-electrons (PE) and randomly accepted accounting for a quantum detection efficiency of 35$\%$. This produces a light yield of about 50~PE/MeV, with a smearing of about $10\%$. To account for a realistic detection threshold hits below 3~PE are rejected~\cite{Blondel:2020hml}.
 \rvwA{As the Bragg peak would be clearly visible even with Birks saturation~\cite{Blondel:2020hml} no quenching corrections have been considered}. Neither attenuation nor bar-to-bar optical crosstalk are included in the ASTRA simulation. This assumptions are reasonable as the light yield in optical crosstalk hits is much lower than in the main hits~\cite{Blondel:2020hml} and therefore can be easily identified and removed using simple light yield cuts or machine learning techniques~\cite{Alonso-Monsalve:2020nde}. The attenuation could be corrected following similar prescriptions to those used in SuperFGD~\cite{Blondel:2020hml}. 
The output of the simulation has a structure that mimics that of the conceptual detector. For a given event
a list of DMAPS hits and ASTRA hits is provided. 

\color{black}
\subsection{Tracking}
Custom algorithms were used to associate hits into tracks. \rvwA{The main goal of this algorithms was to provide a sufficient performance to evaluate the potential of the simulated pCT system. Although, in principle, further optimizations with superior medical and computational performance might be achieved by improving the reconstruction software, such a dedicated task is out of the scope of this studies, despite it might be addressed in the future as a natural continuation to this work. Accordingly, it is worth noting that the figures presented in the results section, specially regarding multi-proton reconstruction, constitute a lower bound of the potential performance of the system.}
\label{sec:tracking_alg}
\subsubsection{DMAPS Tracking}
\begin{figure}[ht!]
    \centering
    \includegraphics[width=0.99\linewidth]{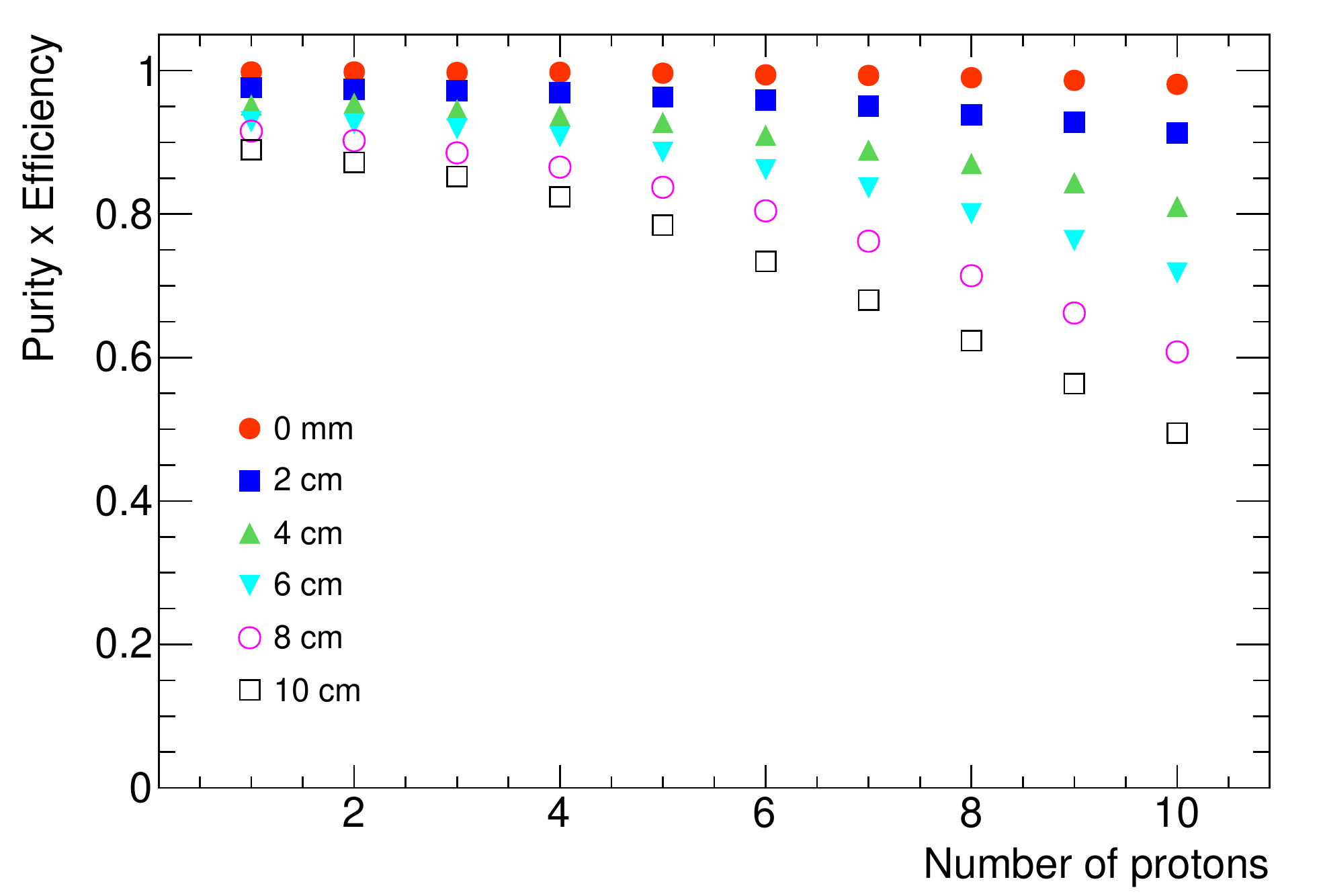}
    \caption{Plot of the purity$\times$efficiency of the DMAPS tracker as function of the number of protons per event for different thicknesses of the phantom in place. The studies are made using a Gaussian beam ($\sigma=10$~mm) containing 180 MeV protons.}
    \label{fig:efftpurity}
\end{figure}
For a given collection of fired pixels in one event, the DMAPS tracking algorithm finds a set of reconstructed trajectories running the following steps:
\begin{itemize}
    \item Define the number of tracks $N$ as the lowest number of hits in a plane.
    \item Generate all the possible track combinations that do not share any commont point and compute a fitness value ($\mathcal{K}$) for each.
    \item Select the set of N tracks that maximize the total fitness.
\end{itemize}
\noindent Two different approaches were explored for the $\mathcal{K}$ parameter. On one hand it was defined as the $\chi^2$ of a straight line fit to all four pixel positions. On the other hand, as the minimum line-to-line distance using the two trajectories reconstructed with the two first and the two last planes. The latter was used as it works better for events with several simultaneous protons. Performance results are presented in Figure~\ref{fig:efftpurity}.
In order to study the performance of the algorithm, two figures of merit were considered, the purity ($p$) and the efficiency ($\varepsilon$), defined as:
\begin{equation}
\varepsilon= \frac{N_{\textup{reconstructed}}}{N_{\textup{total}}},\qquad p =\frac{N_{\textup{good}}}{N_{\textup{reconstructed}}}
\label{eff}
\end{equation}
where $N_{\textup{total}}$, $N_{\textup{reconstructed}}$ and $N_{\textup{good}}$ stand respectively for the total number of simulated tracks, the total number of reconstructed tracks, and the total number of reconstructed tracks with all hits belonging to the same true track.
 
\subsubsection{ASTRA Tracking}

\begin{figure}[ht!]
\centering
\begin{minipage}{0.95\linewidth}
    \centering
    \includegraphics[width=0.99\linewidth]{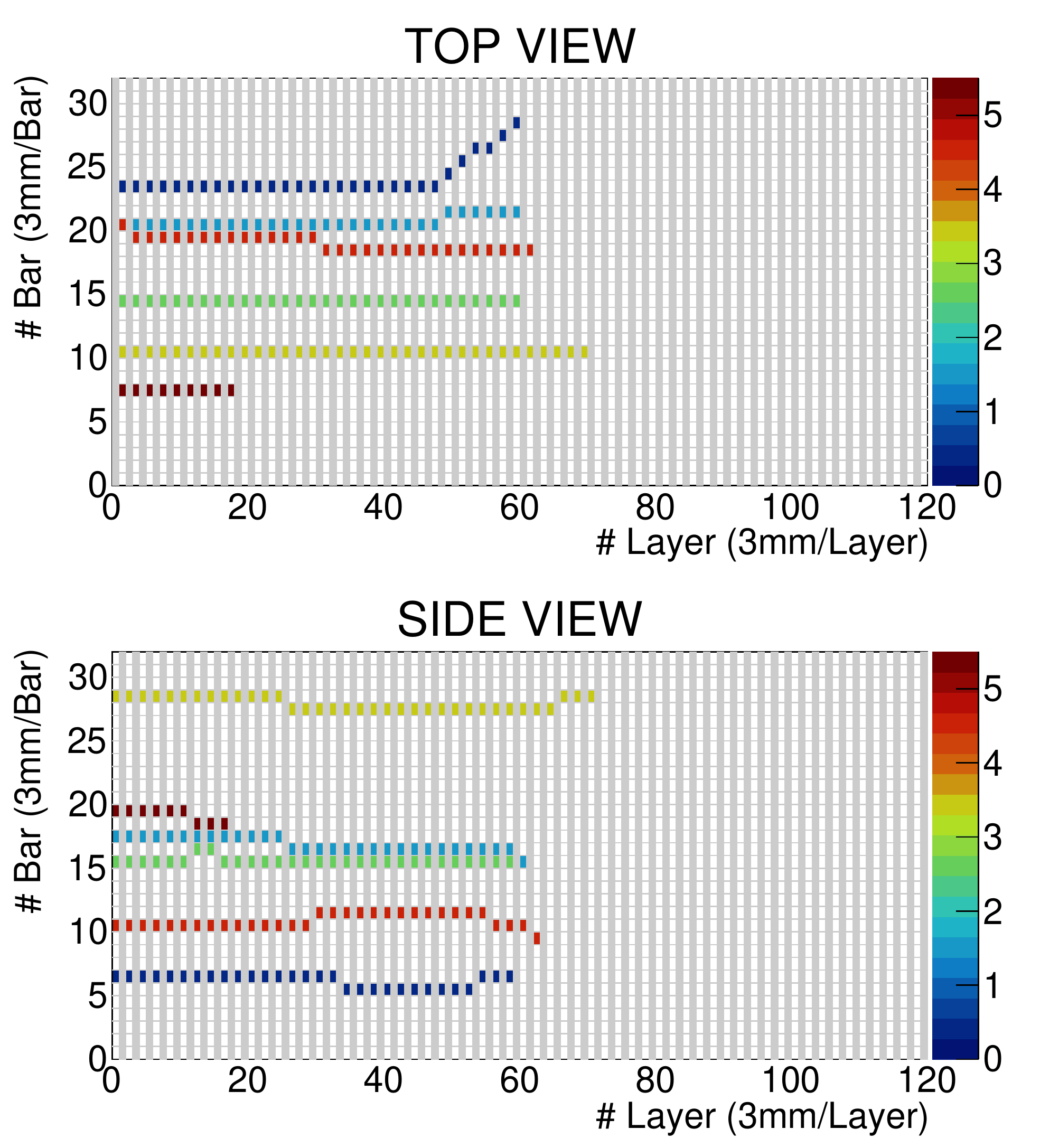}
    \includegraphics[width=0.99\linewidth]{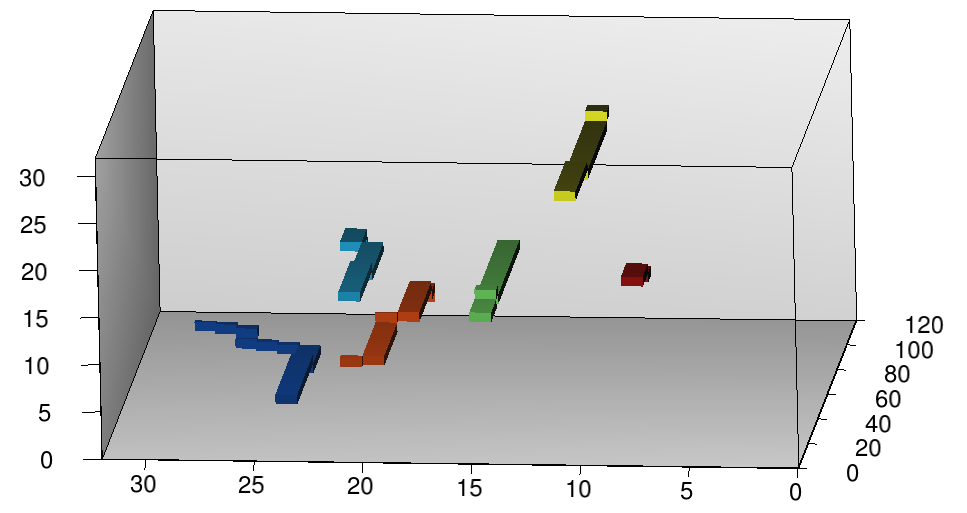}
\end{minipage} 
\caption{Reconstructed event with six simultaneous protons in ASTRA. The displays above represent the 2D hits in the top and side views respectively. The bottom display is a 3D representation of the reconstructed tracks. Each color represents a different reconstructed track ID.}
\label{fig:astra_detail}
\end{figure}
\noindent For a given collection of bar hits in one event, the ASTRA tracking algorithm finds a set of reconstructed trajectories running the following steps:
\begin{itemize}
    \item Make all possible 3D point combinations using the two first ASTRA layers. 
    \item Set as track seeds all 3D points closer than a distance $D$ to the trajectory defined by the last two DMAPS.
    \item For each seed, it iterates going upstream layer by layer. For each new layer, 3D point candidates are formed from the available hits. All candidates must be closer than a distance $D$. The closest candidate is added to the track.
   \item If no new candidates are found a new reconstructed track is formed. The hits used for the track are set as unavailable and the algorithm continues with the next seed until all seeds are processed.
\end{itemize}
An example of the tracking result of an event containing six simultaneous protons is presented in Figure~\ref{fig:astra_detail}.

\subsection{WEPL Calibration}
\label{sec:WaterTankTest}
\begin{figure}[ht!]
    \centering
    \includegraphics[width=0.92\linewidth]{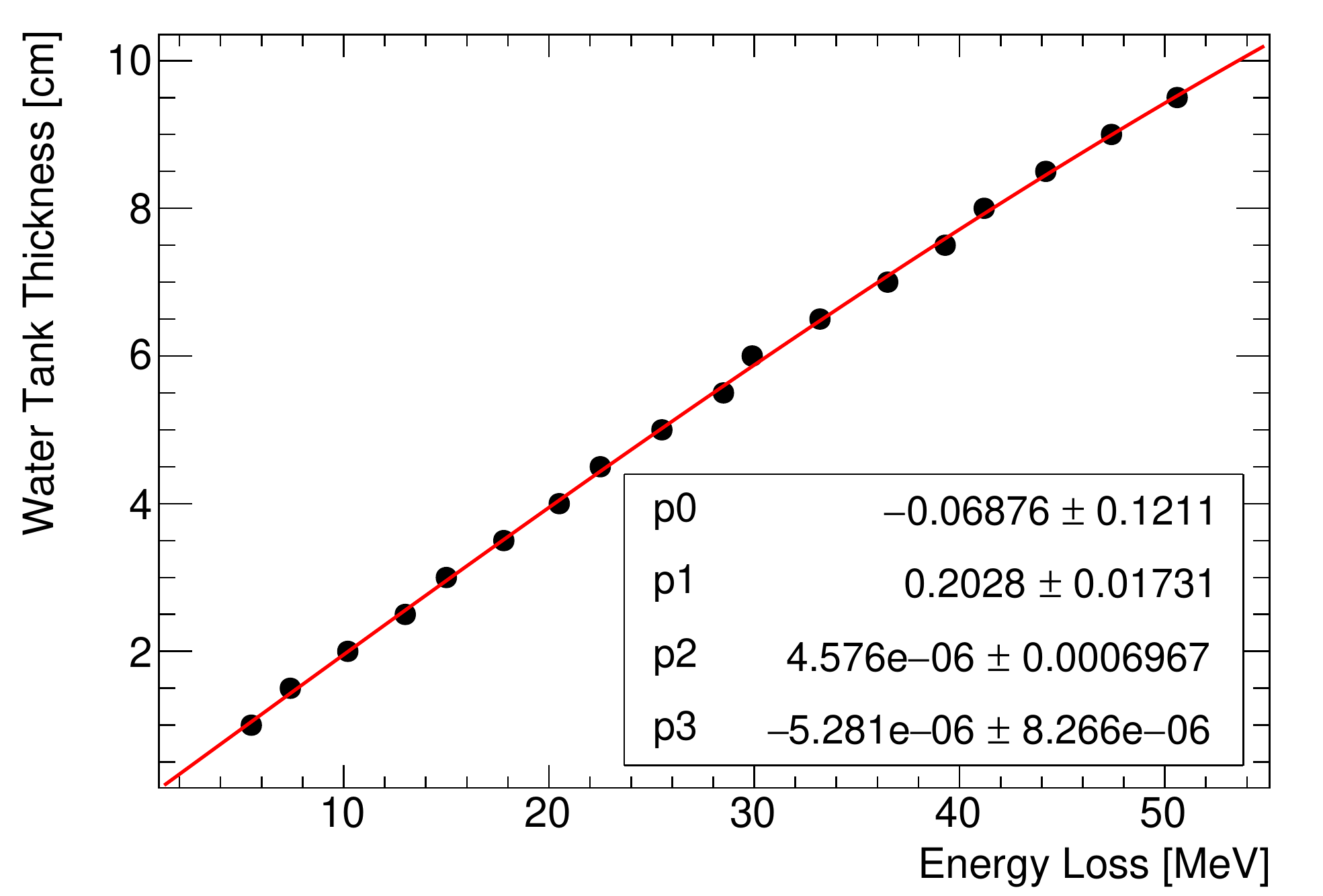}
    \caption{Water tank thickness as function of the energy loss for a monochromatic 180~MeV proton beam. The simulated data is fit using a 3rd degree polynomial used to compute the WEPL on the radiography of the squared phantom.}
    \label{fig:WTTest}
\end{figure}

To convert the energy loss into Water Equivalent Path Length (WEPL) units a Water Tank (WT) test calibration was made. The results, presented in Figure \ref{fig:WTTest}, show the WT thickness as function of the energy loss for a monochromatic 180~MeV proton beam. 

\subsection{Performance Tests}
To study the imaging capabilities of the system, a series of performance tests with phantoms placed between the second and third DMAPS were made. Unless otherwise specified a 180~MeV monoenergetic proton beam with a Gaussian profile ($\sigma=10$~mm) was used \rvwA{ well matching the characteristics of the iThemba proton beam facility~\cite{esposito2015cmos}.}

\subsubsection{Energy Reconstruction by Range}
\begin{figure}[ht!]
    \centering
    \includegraphics[width=0.99\linewidth]{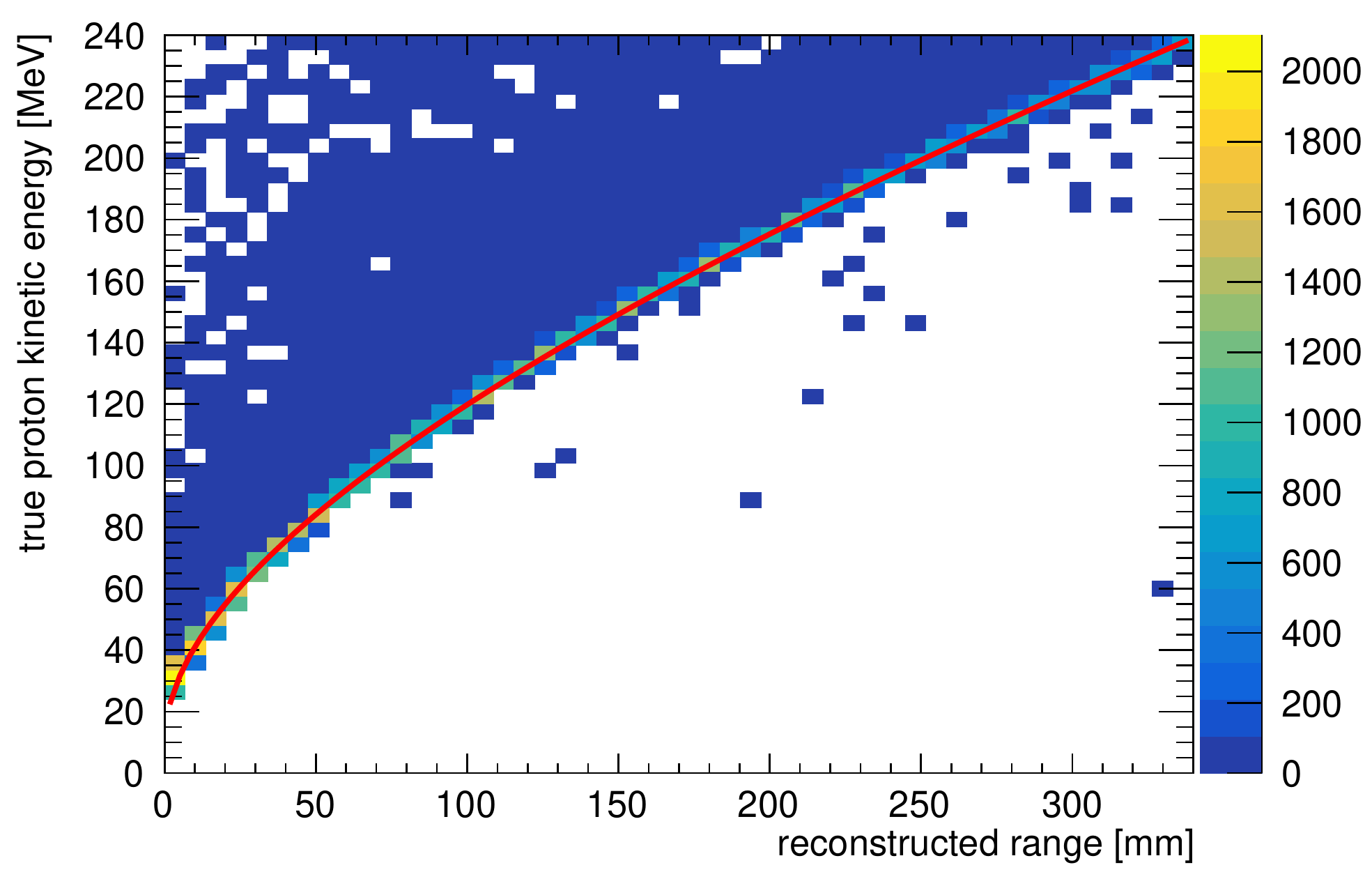}
    \caption{True proton kinetic energy from GEANT4 compared to the reconstructed range in ASTRA. The map from the reconstructed range to the reconstructed energy corresponds to the fit in red.}
    \label{fig:range_to_e_map}
\end{figure}

\begin{figure}[ht!]
    \centering
    \includegraphics[width=0.99\linewidth]{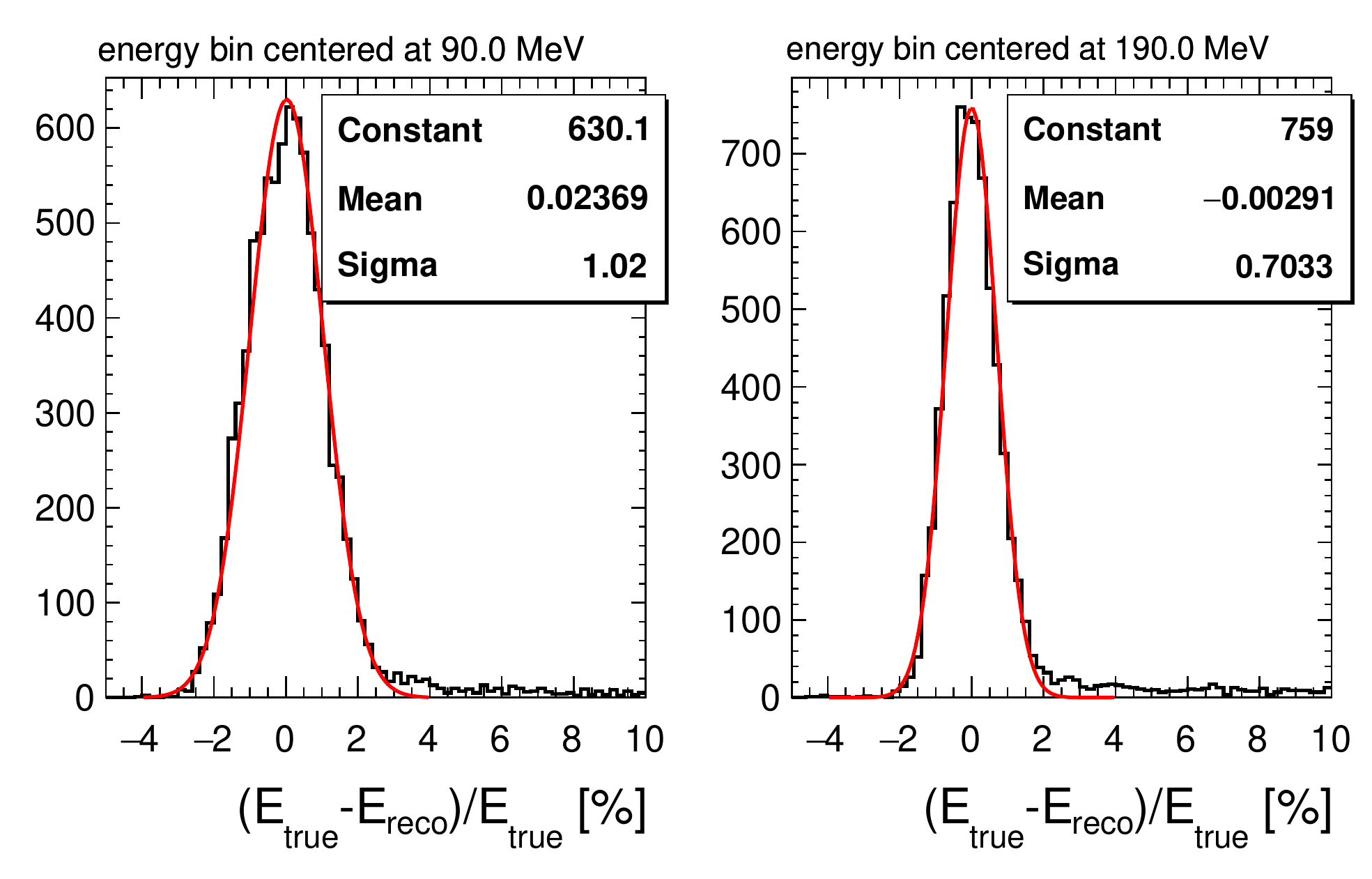}
    \caption{Examples of two of the distributions for bars of 3~mm used in Figure~\ref{fig:eresbythickness}. The left (right) plot corresponds to a resolution of 1.02$\%$ (0.70$\%$).}
    \label{fig:fit_bins_eres}
\end{figure}
\label{sec:e_recon}
In order to reconstruct the protons kinetic energy by range using ASTRA the strategy was to build a map from the reconstructed range in ASTRA to the true kinetic energy of the proton provided by GEANT4, as presented in Figure~\ref{fig:range_to_e_map}. The map from the reconstructed range to the reconstructed energy is obtained fitting the most likely true energy for a given range. To compute the energy resolution, as presented in Figure~\ref{fig:eresbythickness}, distributions of $1-\textup{E}_{\textup{true}}/\textup{E}_{\textup{reco}}$ were filled for all reconstructed protons. The distributions covered intervals of 20~MeV. Each of the distributions was then fit using a Gaussian function and the sigma of the fit was used as the energy resolution. Examples of two of this distributions and fits are presented in Figure~\ref{fig:fit_bins_eres}.

\subsubsection{Energy Reconstruction Including Calorimetry}
The potential of including calorimetric information from ASTRA, in addition to the range, to reconstruct the protons energy was studied. In order to do so, the reconstructed range and the light yield from every hit associated to each reconstructed proton was used to train a boost decision tree (BDT). In particular, the BDTG method from the TMVA libraries~\cite{Therhaag:2009dp} has been used. Half a million reconstructed protons have been used for training. Later, the trained BDT was used to predict the reconstructed energy with independent data, namely, events not used to train the algorithm.

\color{black}
\subsubsection{Imaging}
\label{sec:imaging}
Imaging tests were performed using the simulated pCT system. Such tests consisted on imaging phantoms, composed by up to seven different materials with densities defined in Table \ref{tab:density}, placed between the second and third DMAPS planes. Two different types of images were performed. A simple 2D radiography and a 3D pCT scan.
\begin{table}[ht!]
    \centering
    \begin{tabular}{|c|c|}\hline
      Material   & Density [g/cm$^3$] \\\hline
        Water & 1.00  \\
       Adipose & 0.92  \\
       Perspex & 1.177 \\
        Lung  & 0.30  \\
        HC bone  & 1.84  \\
        Rib bone  & 1.40  \\
        Air  & 1.3$\times 10^{-3}$  \\
        \hline
    \end{tabular}
    \caption{Density values of the simulated materials used for imaging.}
    \label{tab:density}
\end{table}
For the radiography a proton scan on the phantom was made moving the center of the Gaussian beam in a squared grid over the phantom surface. The image coordinates were reconstructed projecting the reconstructed tracks trajectory on an imaginary plane, perpendicular to the beam, located at the center of the phantom. A 2D grid of 200$\times$200 image pixels of 400$\times$400~\textmu m$^2$ was defined on that plane, covering a total area of 8$\times$8~cm$^2$. For each grid-pixel the protons reconstructed energy spectrum was stored, \rvwA{without correcting for their path-length in the phantom,} in a 1D histogram with bins of 1~MeV width. Later, a reconstructed energy was associated for each grid-pixel as the mean from a Gaussian fit to its reconstructed energy spectrum. For the pCT image, 360 radiography images were used, rotating by one degree the phantom for each scan. We accepted as protons good for imaging all reconstructed tracks with a reconstructed energy in a $2~\sigma$ range around most probable energy on its corresponding grid-pixel in the associated 2D radiography. This value was chosen as it accepted most of the protons good for imaging while removing possible tails. The motivations for this are further discussed in the next section. 
\rvwA{The position and direction of the accepted protons at each plane, and the reconstructed energy of each proton were used as inputs to an algorithm developed by the PRaVDA collaboration \cite{ESPOSITO2018149} which outputs RSP tomographic images. For the 2D images, the water tank calibration was only used to convert the energy loss into WEPL.} All images were made using energy reconstructed exclusively by range, without considering any calorimetric information.

\section{\rvwA{Results and Discussion}}\label{sec:Results}
\begin{figure}[ht!]
	\centering
	\includegraphics[width=0.99\linewidth]{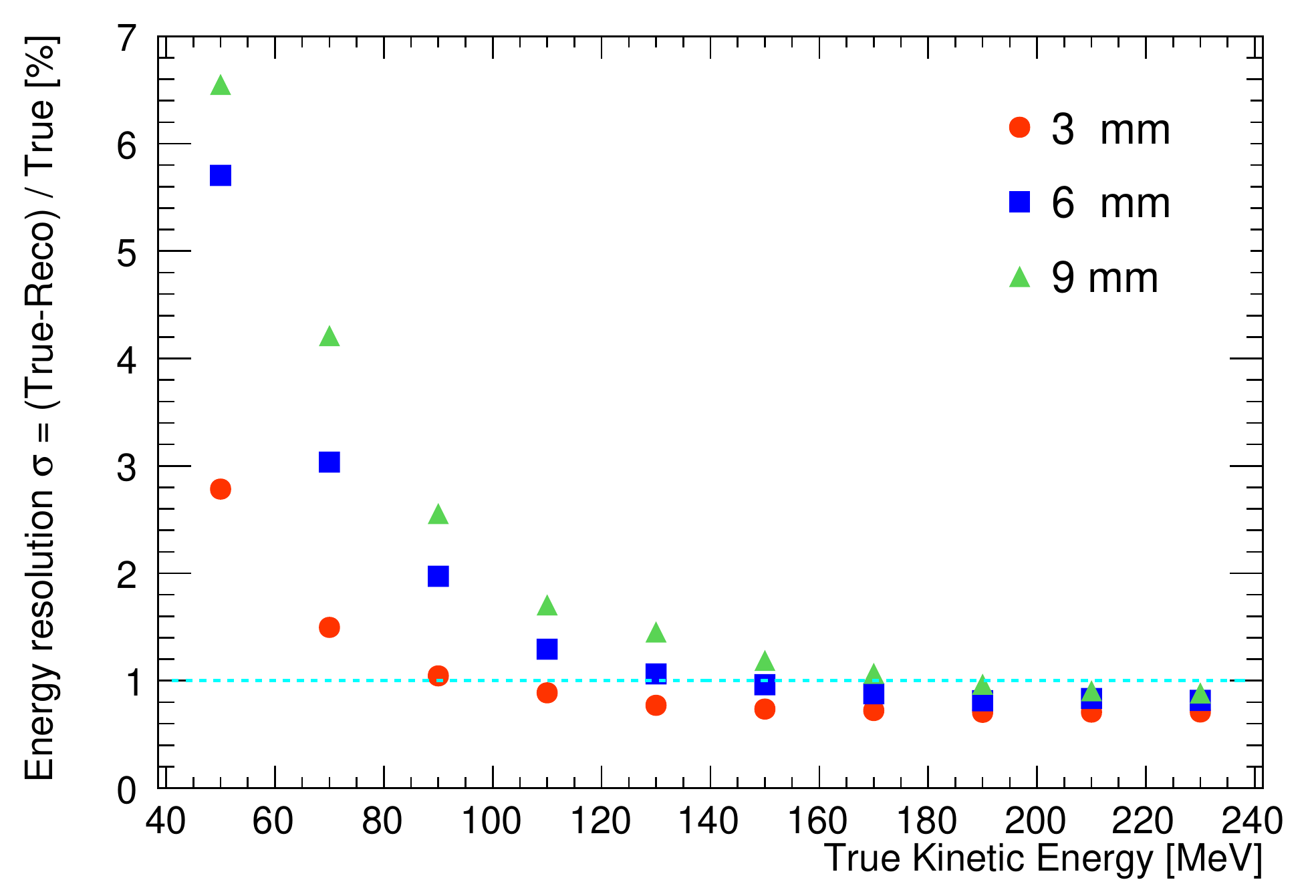}
	\caption{Energy resolution of the ASTRA range telescope using range only information for three different squared-shaped bar sizes of 3, 6, and 9~mm. The dashed line highlights the 1~$\%$ threshold.}
	\label{fig:eresbythickness}
\end{figure}

\begin{figure}[ht!]
	\centering
	\includegraphics[width=0.99\linewidth]{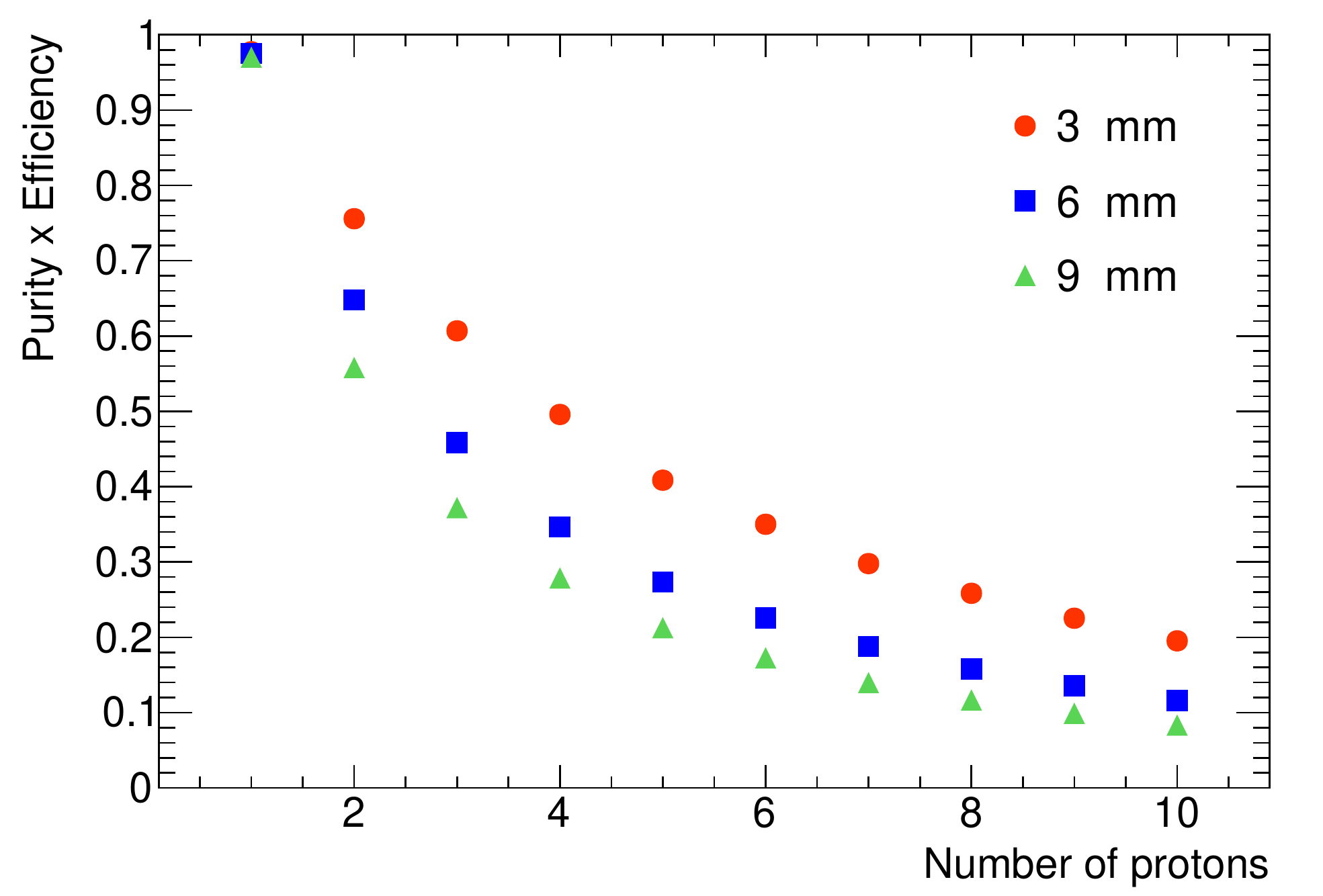}
	\caption{Purity times efficiency of reconstructed tracks \rvwA{in the pCT system (DMPAS+ASTRA)} for different number of simultaneous protons for the 3~mm, 6~mm and 9~mm squared shaped ASTRA bars configurations.}
	\label{fig:pur_thickness}
\end{figure}

\begin{figure}[ht!]
	\centering
	\includegraphics[width=0.99\linewidth]{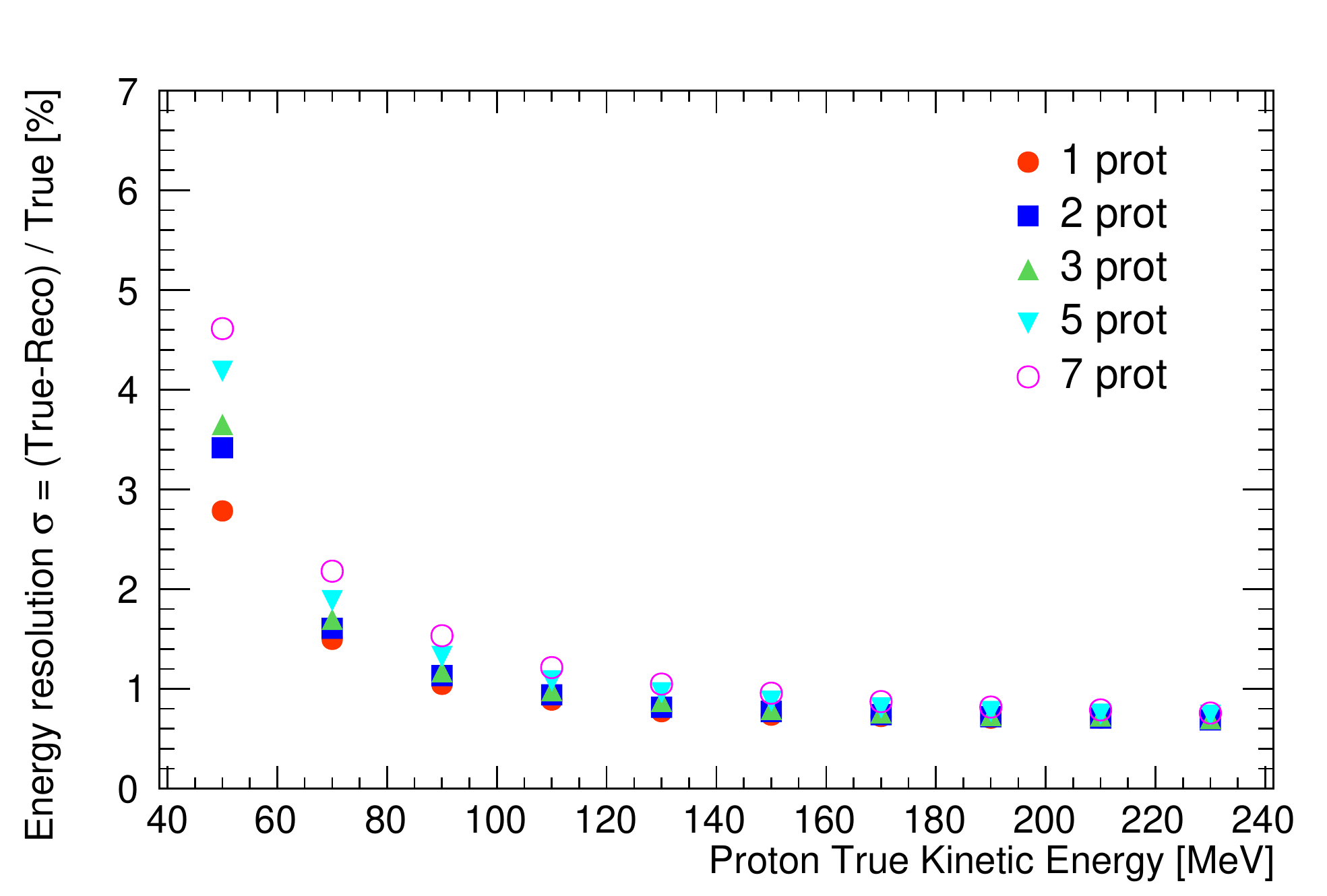}
	\caption{Energy resolution for different number of simultaneous protons as a function of the proton initial kinetic energy. }
	\label{fig:eres_by_multiplicity}
\end{figure}

The energy resolution by range in ASTRA, computed following the details in \ref{sec:e_recon}, is presented for different bar sizes in Figure~\ref{fig:eresbythickness}.
Overall, the system exhibits a sub-1$\%$ energy resolution for the 3~mm configuration for protons with energies above 100~MeV. The energy resolution approaches $\sim$0.7$\%$ asymptotically. A similar performance is achieved for even coarser segmentations in the high energy limit, \rvwA{opening the possibility to reduce the number of channels of ASTRA and therefore its production costs}. Using thick bars has, however, a caveat. As presented in Figure~\ref{fig:pur_thickness}, the multi-proton tracking capacity of \rvwA{the pCT system}, in terms of purity and efficiency, is significantly better for the 3~mm configuration. Hence, \rvwA{to explore this possibility}, for the rest of the studies we focus on 3~mm bars.
\\When reconstructing the energy of protons exclusively by range, some miss-reconstructions are unavoidable. Protons experience inelastic interactions which shorten their expected range contributing to very long tails to the reconstructed energy. In addition, some tracking errors might lead to inaccurate range estimates. Thus, in the reconstructed energy distributions there are two regimes, a Gaussian distributed one arising from the correct reconstruction of elastic protons, and a one conformed by long tails produced by tracking errors and inelastic protons. This features can be seen in Figure~\ref{fig:fit_bins_eres} which shows the percentage error in the reconstructed energy by range. Of course, for events with a single proton tracking errors are expected to be close to zero and the tail contributions to come mainly from inelastic interactions but for an increasing number of simultaneous protons tracking errors are more common, as the energy of the two trajectories, or some of their hits, might be swapped. Thus, a relevant figure of merit to understand the expected performance of the detector is to quantify how many reconstructed protons are good for imaging. For a pCT system with multi-proton capabilities this depends significantly on the beam profile: the narrower the beam, the harder to correctly identify the hits associated to each proton.
\begin{figure}[ht!]
	\centering
	\includegraphics[width=0.99\linewidth]{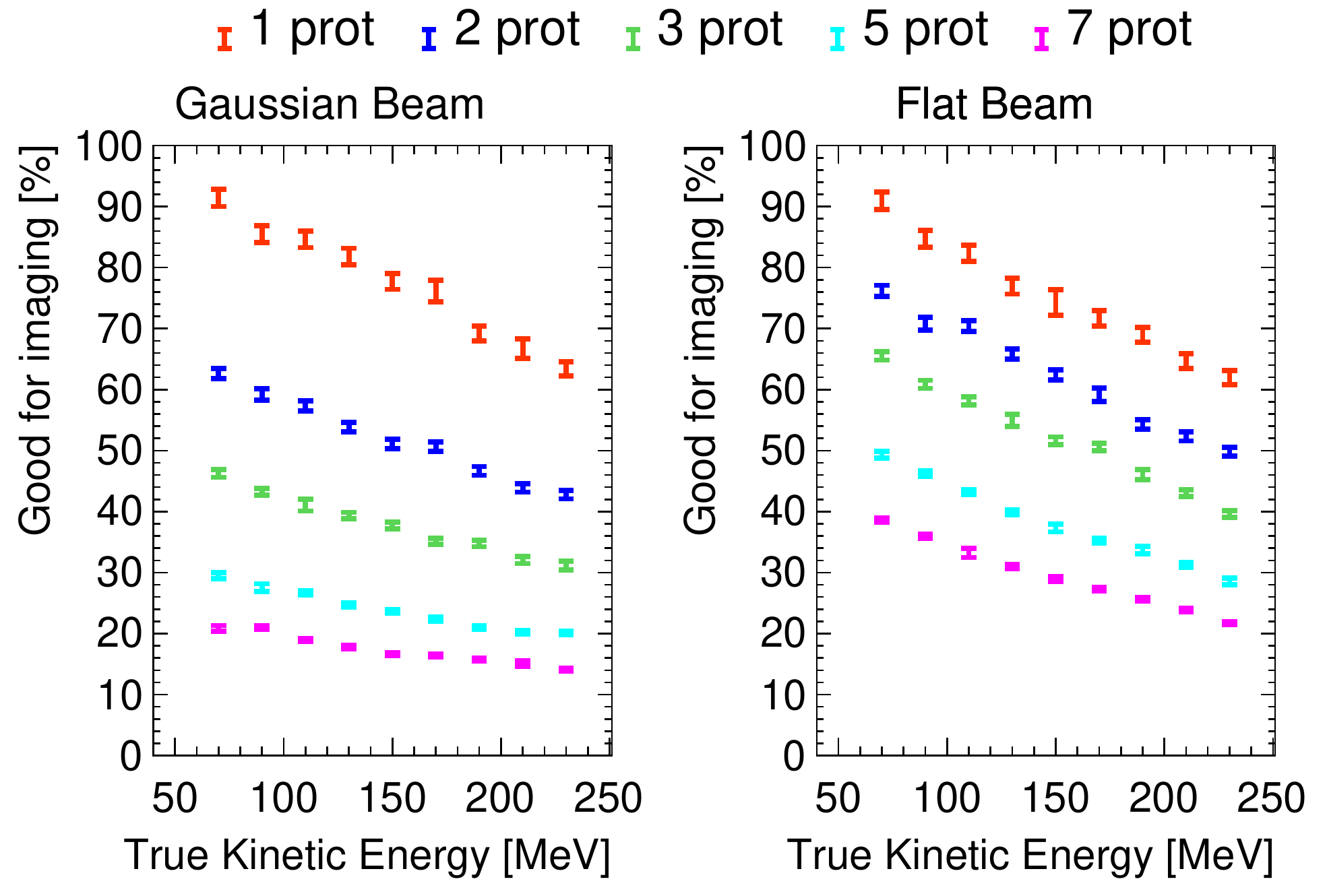}
	\caption{Fraction of protons good for imaging as a function of the proton true kinetic energy using different number of simultaneous protons for a Gaussian beam ($\sigma=10~mm$) and for a flat 75$\times$75~mm$^2$ beam.}
	\label{fig:frac_gauss_prots}
\end{figure}
\\To illustrate this dependence, we present in Figure~\ref{fig:frac_gauss_prots} the fraction of protons good for imaging as a function of the number of simultaneous protons for two different beam profiles. The fraction of protons good for imaging, is obtained fitting each distribution of energy resolutions with a Gaussian and counting the fraction of events within 2$\sigma$ with respect to the total number of incident protons, such that the pCT system efficiency is considered implicitly. \rvwA{Remarkably, as earlier presented in Figure~\ref{fig:efftpurity}, the limiting factor for multi-proton tracking for the pCT system under consideration would come from the segmentation in ASTRA, given that the fine pixelization of the DMAPS tracker allows to efficiently identify multiple protons at the same time with excellent purity.} As expected, \rvwA{the results for ASTRA show that} the fraction of good protons is higher if the simultaneous protons are typically more spaced. Remarkably, even for the most challenging of the two beam profiles, which corresponds to a realistic clinical beam, about 1 proton per bunch can be used for imaging regardless of whether there is one, two or three simultaneous protons in the bunch. For the wider beam, even for the bunches with 3 and 5 protons about half of them are good for imaging. Finally, notice that, independently of the beam configuration, for higher energies the range is longer and the probability of experiencing an inelastic interaction grows, reducing the fraction of reconstructed protons which are good for imaging. In addition, one might wonder if the width of the sigma defining the Gaussian regime increases with the number of protons. This, however, has a very small effect as presented in Figure~\ref{fig:eres_by_multiplicity}.\\
In a real pCT system the true proton energy is not known such that identifying which protons are good for imaging is not straightforward. A common solution is to build a classifier which tries to identify protons bad for imaging, e.g. searching for kinks or a missing Bragg peak in the energy tagger trajectory. The remaining protons are label as good. Recently, authors have reported a $\sim 97\%$ accuracy in this task using a CNN~\cite{pettersen2021investigating}. This method, however, is not entirely satisfactory for a multi-proton pCT system as sometimes the correctly reconstructed energy of two simultaneous protons is swapped due to matching errors between the position tracker and the energy tagger. To overcome this, we use the grid method detailed in section~\ref{sec:imaging} to determine which protons are good for imaging. Conceptually, the method consists in estimating the true energy in each pixel of the grid as the most probable value in the pixel’s distribution and use the distance between the expected and reconstructed energies to set up a selection criteria.\\

\begin{figure}[ht!]
	\centering
	\includegraphics[width=0.99\linewidth]{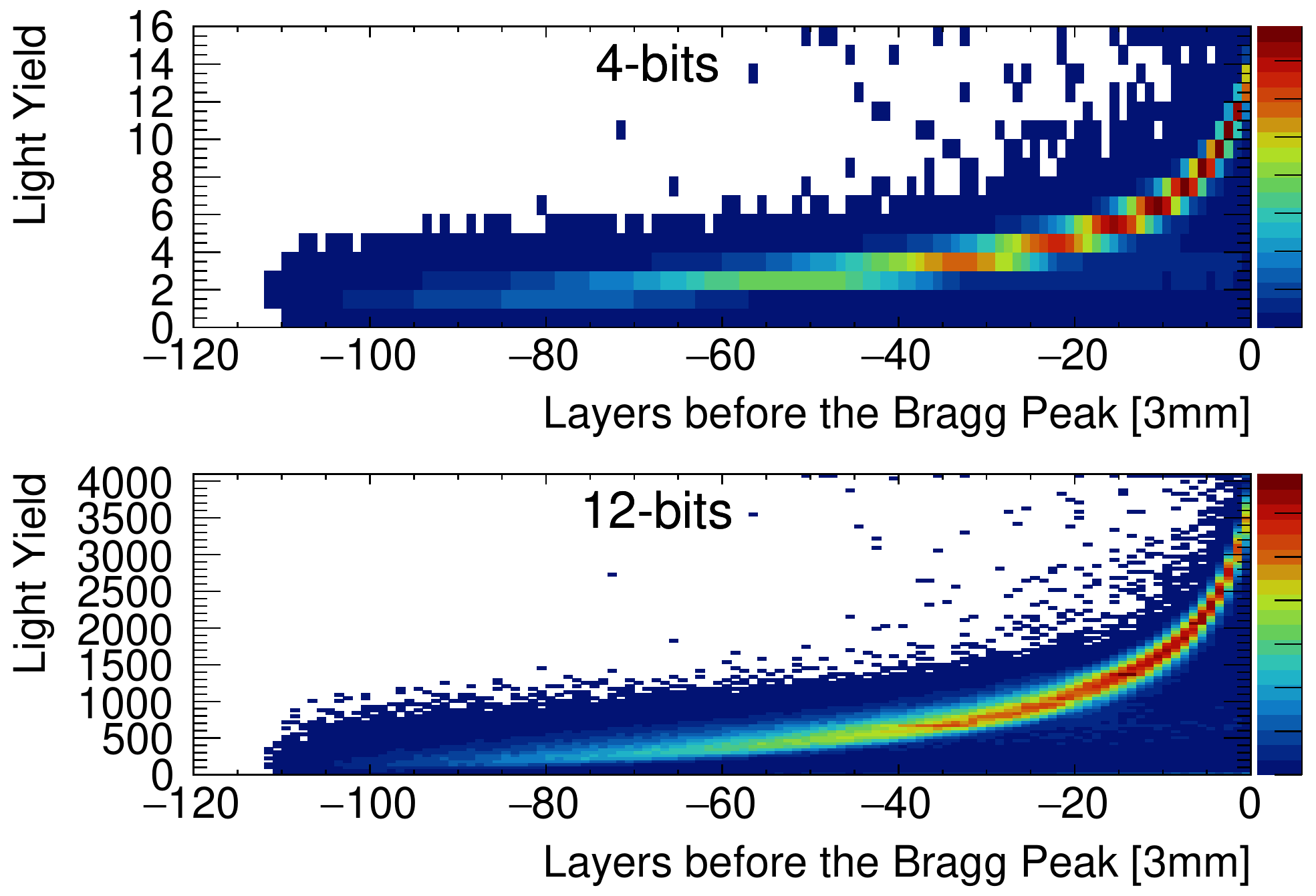}
	\caption{Reconstructed protons light yield as a function of the distance to the layer with maximum recorded light yield using the 3~mm bar configuration and two different ranges of values. The initial protons energy is flat in the range of 40 to 240~MeV.}
	\label{fig:bragg_ADC}
\end{figure}

\begin{figure}[ht!]
	\centering
	\includegraphics[width=0.99\linewidth]{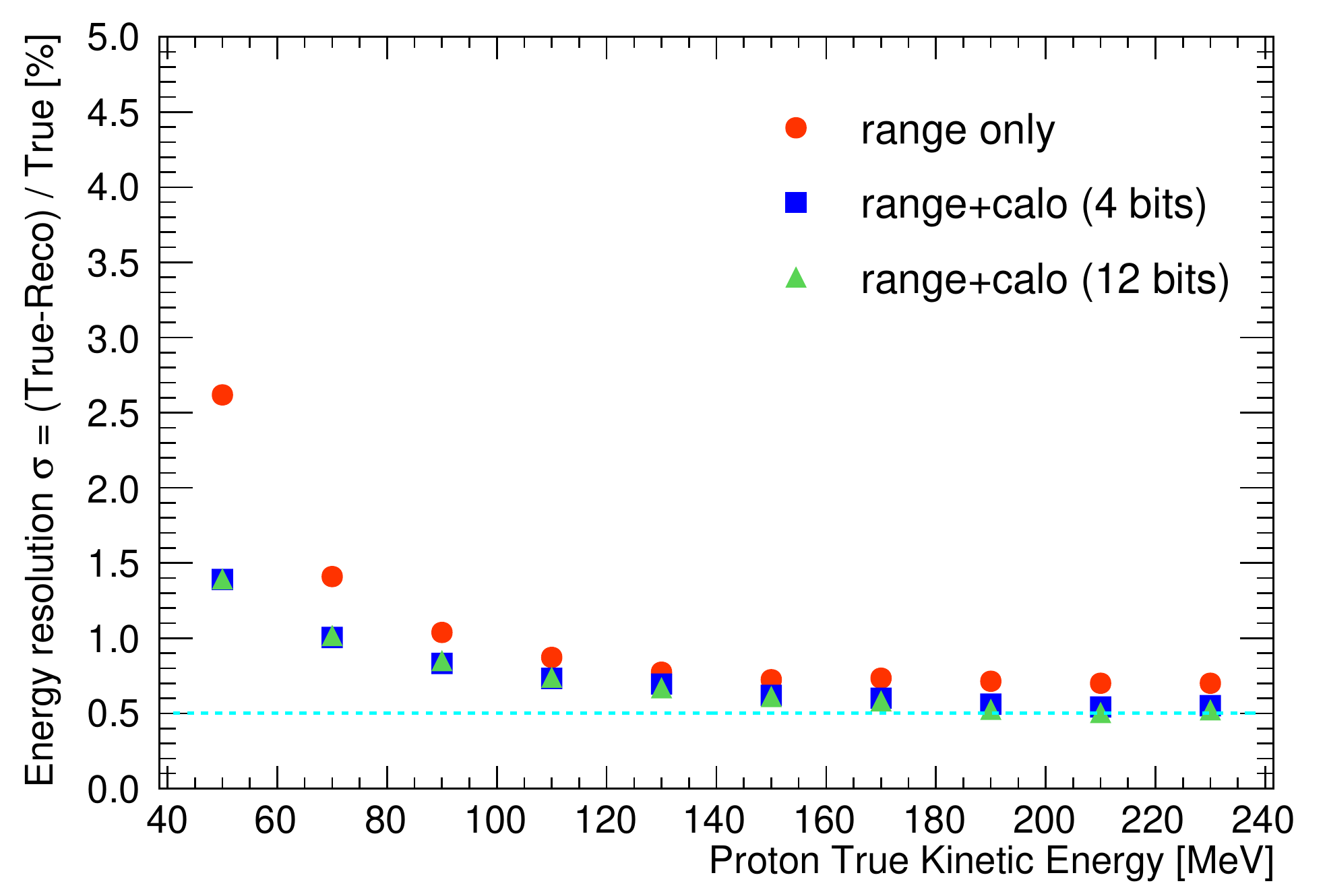}
	\caption{Energy resolution for single proton events with and without using calorimetric information. For the later two configurations are tested, one discretizing the light yield in 16 values (4-bits ADC), and another discretizing the light yield in 4096 values (12-bits ADC).  The dashed line highlights the 0.5~$\%$ threshold.}
	\label{fig:eres_range_calo}
\end{figure}

In order to study the benefits of using calorimetric information in ASTRA, we compare the range only approach to an energy reconstruction method combining range and calorimetry. Considering the high data rate to produce the pCT, using small data transfers is preferred. Thus, we simulated the light yield as being discretized in 2$^N$ values, to account for the impact of using an N-bits ADC. Two configurations have been considered, one with 4-bits (16 values), and one with 12-bits (4096 values). The difference in the quality of the calorimetric information for each configuration can be seen in Figure~\ref{fig:bragg_ADC}.
The associated performance for the energy resolution is presented in Figure~\ref{fig:eres_range_calo}. The results show a significant improvement for low proton kinetic energies. At high energies the energy resolution improves from $\sim0.7\%$ to $\sim0.5\%$. The results show that using a 12-bit ADC does not provide further performance benefits compared to a 4-bit ADC.
\subsection{Radiography}
\begin{figure}[ht!]
	\centering
	\includegraphics[width=0.89\linewidth]{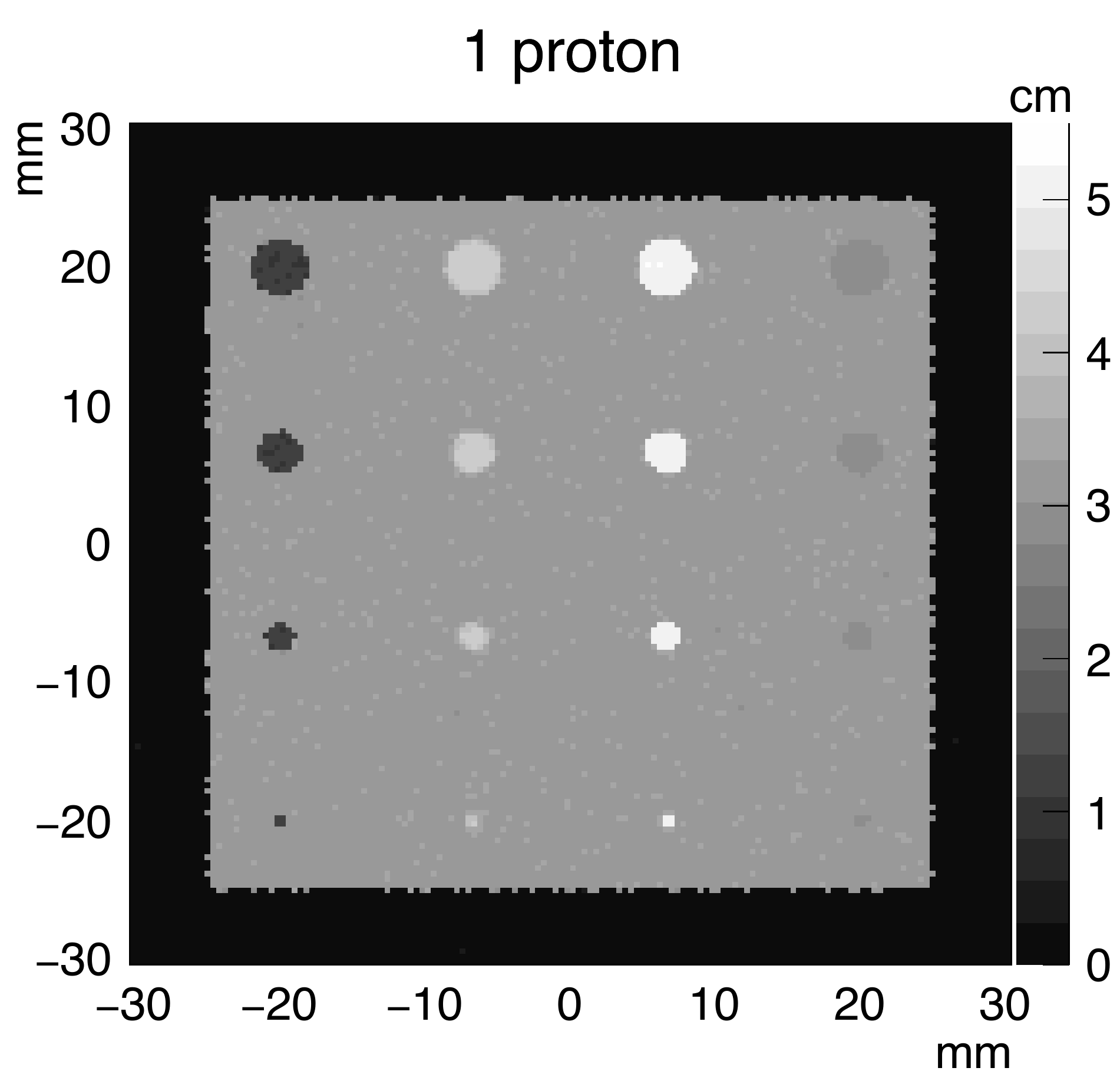}
	\includegraphics[width=0.89\linewidth]{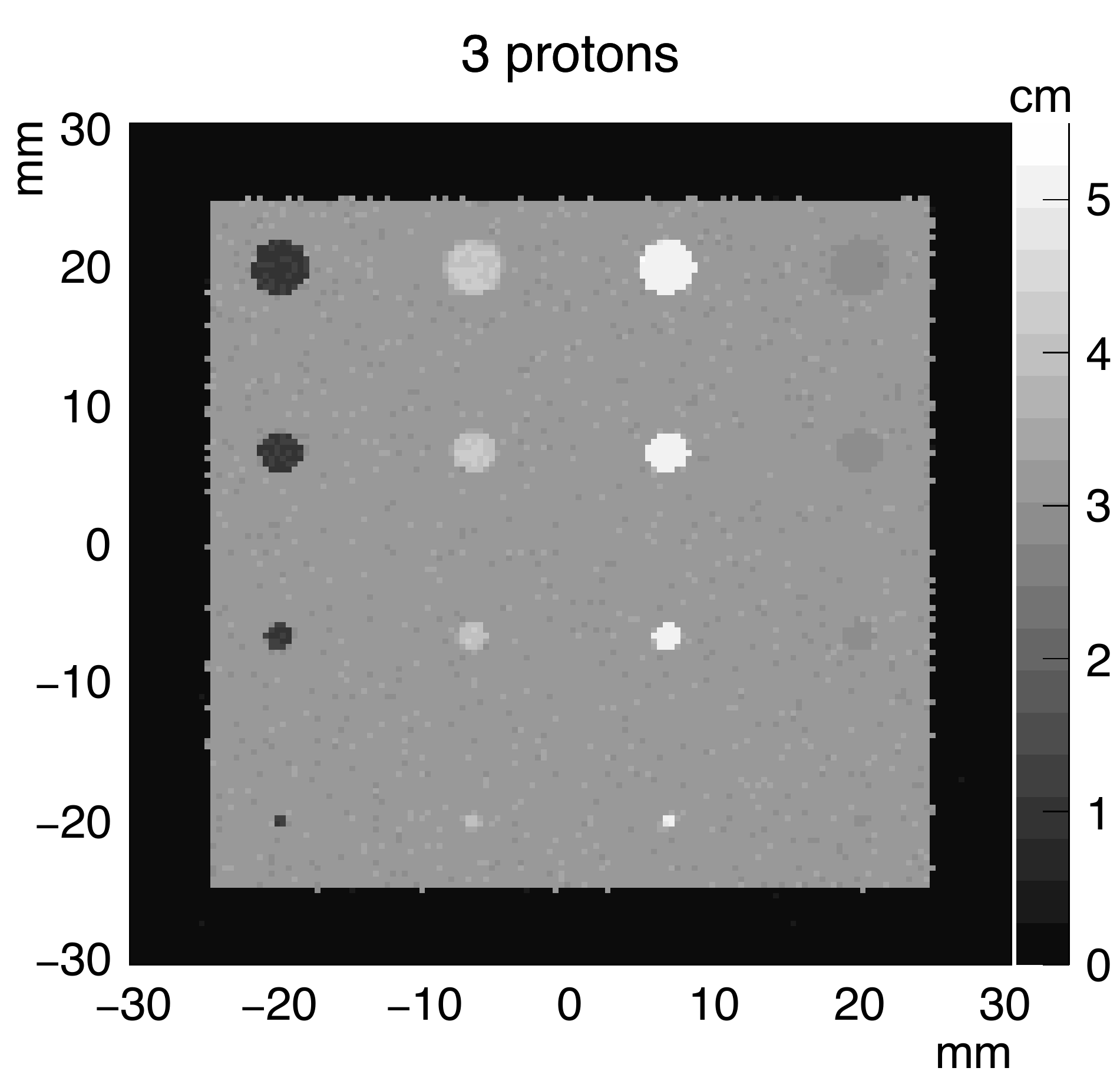}
	\caption{Proton radiography of the squared phantom using 1 and 3 protons. From left to right the materials are simulated as equivalent to lung tissue, rib bone, hard cortical bone, and adipose tissue. From the bottom to the top row the radius are 0.5, 1.0, 1.5 and 2~mm. Each image uses 5$\cdot10^6$ protons. The Z-axis (color) corresponds to Water Equivalent Path Lenght (WEPL) in mm and has been obtained using the energy loss and the data in~\ref{sec:WaterTankTest}.}
	\label{fig:radiography}
\end{figure}

A radiography has been performed, using events with a single proton and three simultaneous protons, on a phantom formed by a simulated water equivalent material (WEM) squared frame of 50x50~mm$^2$ and 30~mm pierced by four columns of cylindrical inserts of 30~mm length. Each column consists of four cylinders of the same material organized in four rows, each with a different radius. From left to right the materials are simulated as equivalent to lung tissue, rib bone, hard cortical bone, and adipose tissue. From the bottom to the top row the radius are 0.5, 1.0, 1.5 and 2~mm. The results are presented in Figure~\ref{fig:radiography}. For all materials and radius the insert leaves a clear signature in the image. Notice that the smaller radius is comparable to the image pixel size of 400$\times$400~\textmu m. As observed, performing the image exclusively with events with 1 proton or with 3 simultaneous protons does not change appreciably the result.

\subsection{Proton Computed Tomography}\label{sec:results_pCT}
The phantom used for the 3D pCT was a spherical phantom that consists in a 75~mm diameter sphere made of Perspex (PMMA) with six different cylindrical inserts 15~mm high with 15mm diameter. The cylinders are placed in a three by three disposition forming two equilateral triangle in two different planes placed 9~mm above and below the center.

\begin{figure}[ht!]
	\centering
	\includegraphics[width=0.89\linewidth]{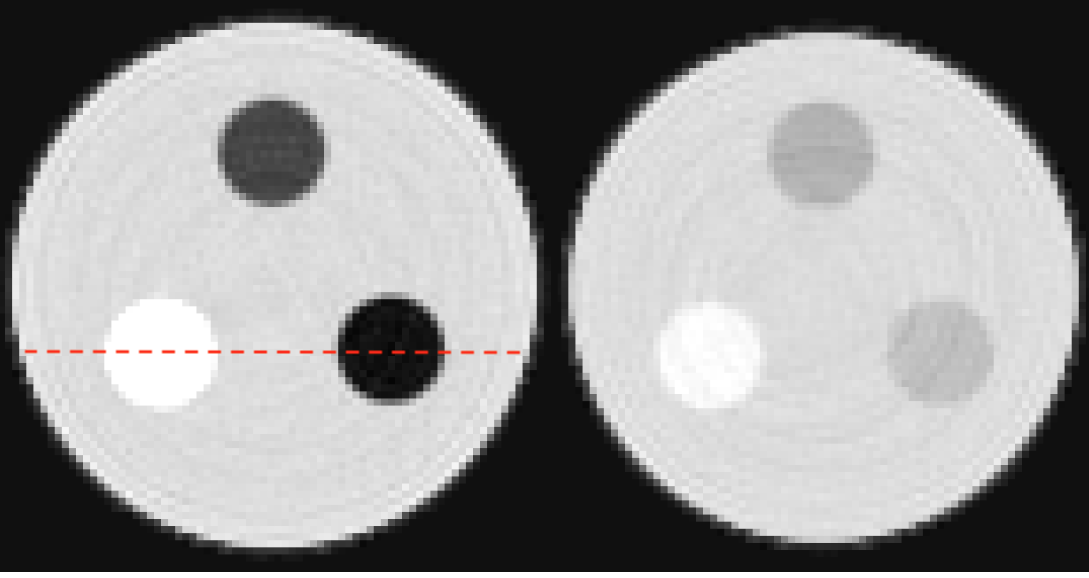}
	\caption{Slices of a proton computed tomography using single proton events showing the contrast in RSP for the six inserts. The different insert materials have been simulated to be equivalent to \rvwA{(from left to right): hard cortical bone, lung and air (left slice) and rib bone, water and adipose tissue (right slice). The red dashed line highlights the data used in Figure~\ref{fig:Insert_Profile}.}}
	\label{fig:pCT_2SetsInserts}
\end{figure}

Figure \ref{fig:pCT_2SetsInserts} shows two sliced sections of a pCT performed using single proton events. Each slice corresponds to the half height of the top (left image) and bottom (right image) sets of inserts. The measured mean values of the RSP for each insert have been computed by selecting the voxel values within the half diameter of each cylinder at six different layers around the center. An equivalent region has been selected to compute the RSP of the perspex frame. The RSP values extracted from Figure \ref{fig:pCT_2SetsInserts} are presented in Table \ref{tab:pCT}. True values have been computed using only true tracks and the true energy of the protons after passing through the phantom in order to provide a reference of the performance. All the reconstructed RSP values, except air, match the reference RSP values within 0.5~$\%$. The air value shows a larger relative discrepancy due to the small RSP of air, comparable to the measurement uncertainty.

\begin{table}[ht!]
    \centering
    \begin{tabular}{|c|c|c|c|c|}\hline
      Material   & RSP (Reco) & RSP (True) & $\%$diff\\\hline
        Water & 0.992 $\pm$ 0.002  & 0.994 $\pm$ 0.002 & 0.201 \\ 
        Air & 0.009 $\pm$ 0.002 & 0.008 $\pm$ 0.002  & -12.5\\
        Adipose & 0.916 $\pm$ 0.006 & 0.917 $\pm$ 0.005 &  0.109\\
        Rib bone & 1.325 $\pm$ 0.003 & 1.326 $\pm$ 0.001 & 0.075\\
        HC bone & 1.641 $\pm$ 0.003 & 1.646 $\pm$ 0.002 & 0.304\\
        Perspex & 1.144 $\pm$ 0.004 & 1.149 $\pm$ 0.002 & 0.455\\
        Lung & 0.302 $\pm$ 0.003 & 0.302 $\pm$ 0.002 & 0.000\\
        \hline
    \end{tabular}
    \caption{Relative Stopping Power (RSP) values for seven different materials extracted from the pCT image of the spherical phantom. The labels \textit{True} and \textit{Reco} stand for the energy used to compute the RSP. To help to compare the values the rightmost column shows the relative difference between the columns on the left.}
    \label{tab:pCT}
\end{table}

\begin{figure}[ht!]
	\centering
	\includegraphics[width=0.89\linewidth]{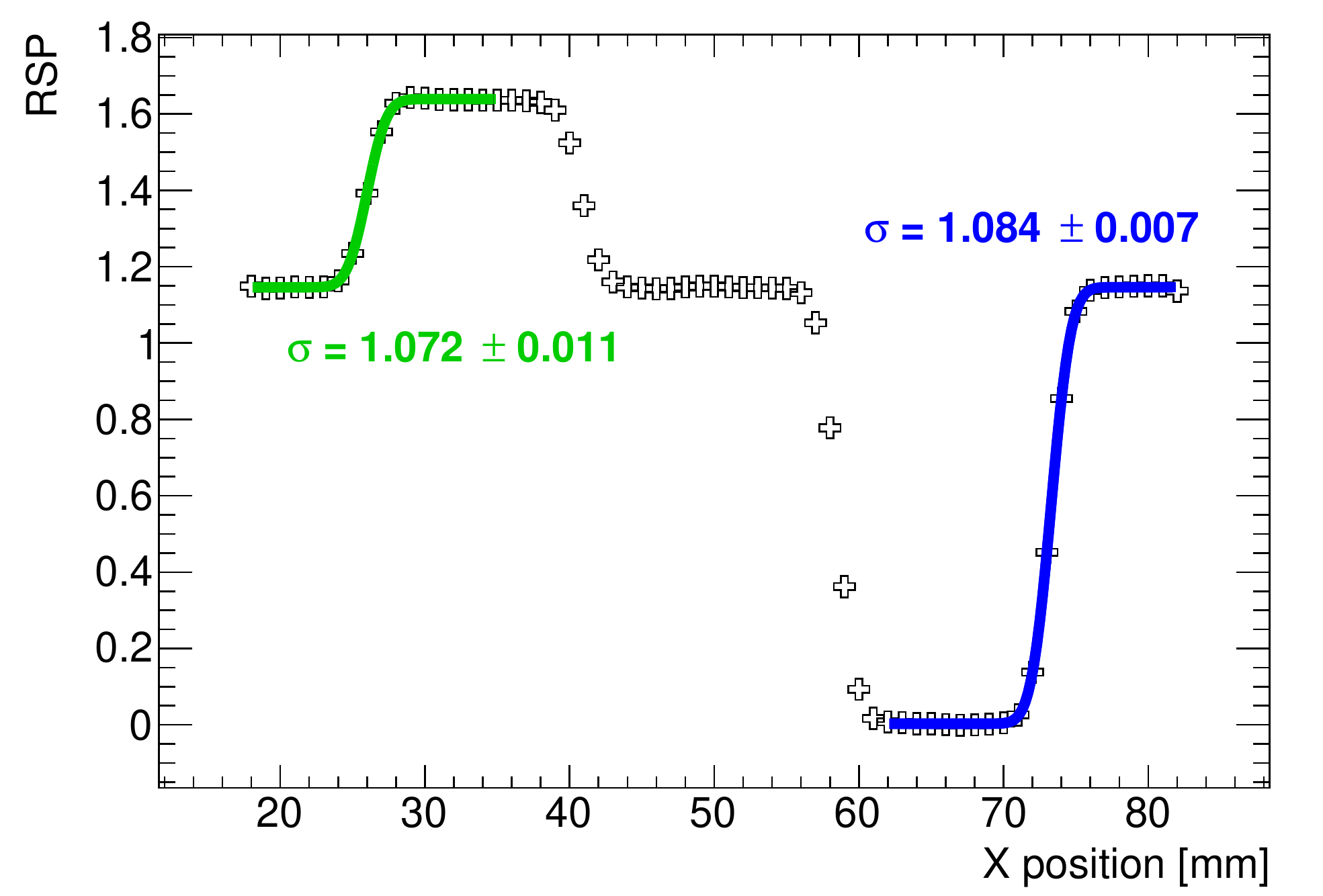}
	\caption{Projection of the RSP along the line highlighted in Figure \ref{fig:pCT_2SetsInserts}. The rise in the RSP value has been fitted with an error function for the two inserts, hard cortical bone (green) and air (blue), characterized by a sigma detailed in the image.}
	\label{fig:Insert_Profile}
\end{figure}
The profile in Figure~\ref{fig:Insert_Profile} shows very stable RSP measurements in the form of smooth trends and flat plateaus. The spatial resolution of the pCT can be characterized by measuring the spread of the transition regions between such plateaus, corresponding to move from outside to inside of an insert (or vice versa). This spread has been quantified to be about 1.1~mm.\\
Finally, the same 3D pCT image has bee made using exclusively events with three simultaneous protons. Following a method analogous to that for the single proton events RSP results are presented in Table~\ref{tab:pCT_3p}. \rvwA{
Although a significant degradation is observed, specially for the materials with lower RSP, the results are of remarkable quality if one considers the fact that only 3-proton events are used. To put it in context, this lower quality results are already competitive with currently existing technologies using only single proton events~\cite{ORDONEZ2017193}. In a real life situation, the beam settings could be configured to ensure a majority of 1-proton events. Multi-proton time frames, often unavoidable due to beam instabilities, would not account for inefficiencies, as in the other existing technologies. Instead, the multi-proton tracking features of ASTRA opens the door to develop reconstruction algorithms that associates different weights to each event depending on its reliability, with the goal to deliver a high quality pCT image. If including multi-proton events to the reconstruction chain would be possible an increased usefulness of the dose delivered to the patient could be achieved paired with a potential reduction of the scan times.}

\begin{table}[ht!]
    \centering
    \begin{tabular}{|c|c|c|c|c|}\hline
      Material   & RSP (Reco 3p) & $\%$diff (True) & $\%$diff (reco 1p)\\\hline
        Water & 1.033 $\pm$ 0.002  &  3.924 & 4.133 \\ 
        Air & 0.076 $\pm$ 0.006 & 850  & 744\\
        Adipose & 0.96 $\pm$ 0.02 &  3.60 &  3.71\\
        Rib bone & 1.34 $\pm$ 0.04 &  1.06 & 1.13\\
        HC bone & 1.66 $\pm$ 0.02 &  0.85 & 1.16\\
        Perspex & 1.14 $\pm$ 0.01 &  -0.78 & -0.35\\
        Lung & 0.35 $\pm$ 0.02 &  15.89 & 15.89\\
        \hline
    \end{tabular}
    \caption{RSP values for the seven different materials of the spherical phantom reconstructed from three proton events compared withe true values and the values reconstructed from single proton events.}
    \label{tab:pCT_3p}
\end{table}

\section{Conclusions}
\rvwA{
In this article a novel range telescope concept is presented based in currently existing and well understood technologies. A full Monte Carlo simulation has been built able to replicate all the relevant features of the proposed system. In its most economic version, with electronics based on single discriminator thresholds, ASTRA would reconstruct the protons energy exclusively by range, with an expected energy resolution similar to 0.7$\%$. The simulations report that more complex electronic choices could improve this value to about 0.5$\%$ by including calorimetric information.  Based on the specifications of the proposed components for ASTRA, an unprecedentedly fast deadtime free proton readout able to cope with 10$^8$ protons/s (100~MHz) is expected. 
To study the imaging performance of the system two phantoms have been used to perform respectively a 2D radiography and a 3D proton computed tomography. The radiography results show excellent detail and contrast even for inserts of sizes comparable to a single pixel. The pCT results show very high 3D contrast, quantified studying the RSP values for different materials, and very good spatial resolution similar to 1.1~mm. Additionally, throughout the paper the additional possibility to use the bar-based ASTRA geometry to reconstruct multi-proton events has been studied. The results highlight the potential of this additional feature which might be exploit in the future to increase the usefulness of the dose delivered to the patient and to reduce the scan times.\\
Overall, the results report great potential for this novel technology which might pave the way towards the development of affordable and high-precision pCT devices capable of working under realistic clinical conditions.}


\section*{Acknowledgments}
\noindent C. Jes\'us-Valls and T. Lux acknowledge funding from the Spanish Ministerio de Econom\'{i}a y Competitividad (SEIDI-MINECO) under Grants No.~PID2019-107564GB-I00 and SEV-2016-0588. IFAE is partially funded by the CERCA program of the Generalitat de Catalunya. F. S\'anchez acknowledge the Swiss National Foundation Grant No. 200021\_85012. This work is supported partly by the EPSRC grant number EP/R023220/1.

\bibliographystyle{ieeetr}
\bibliography{bibliog.bib}

\end{document}